\documentclass[final,12pt]{elsarticle}

\usepackage{graphicx}%
\usepackage{multirow}%
\usepackage{amsmath,amssymb,amsfonts}%
\usepackage[nice]{nicefrac}
\usepackage{amsthm}%
\usepackage{mathrsfs}%
\usepackage{xcolor}%
\usepackage{textcomp}%
\usepackage{manyfoot}%
\usepackage{booktabs}%
\usepackage{algorithm}%
\usepackage{algorithmicx}%
\usepackage{algpseudocode}%
\usepackage{listings}%
\usepackage{subcaption}%
\usepackage{bbm}
\usepackage{array}
\usepackage{url}
\usepackage[normalem]{ulem}
\DeclareMathOperator*{\argmin}{arg\,min}

\begin{document}

\begin{frontmatter}

\title{Few-shot learning for COVID-19 Chest X-Ray Classification with Imbalanced Data: An Inter vs. Intra Domain Study}

\author{Alejandro Galán-Cuenca, Antonio Javier Gallego, Marcelo Saval-Calvo, Antonio Pertusa} 

\affiliation{organization={University Institute for Computer Research},
            city={San Vicente del Raspeig},
            postcode={E-03690}, 
            state={Alicante},
            country={Spain}}

\begin{abstract}
Medical image datasets are essential for training models used in computer-aided diagnosis, treatment planning, and medical research. However, some challenges are associated with these datasets, including variability in data distribution, data scarcity, and transfer learning issues when using models pre-trained from generic images. This work studies the effect of these challenges at the intra- and inter-domain level in few-shot learning scenarios with severe data imbalance. For this, we propose a methodology based on Siamese neural networks in which a series of techniques are integrated to mitigate the effects of data scarcity and distribution imbalance. Specifically, different initialization and data augmentation methods are analyzed, and four adaptations to Siamese networks of solutions to deal with imbalanced data are introduced, including data balancing and weighted loss, both separately and combined, and with a different balance of pairing ratios. Moreover, we also assess the inference process considering four classifiers, namely Histogram, $k$NN, SVM, and Random Forest. 
Evaluation is performed on three chest X-ray datasets with annotated cases of both positive and negative COVID-19 diagnoses. The accuracy of each technique proposed for the Siamese architecture is analyzed separately and their results are compared to those obtained using equivalent methods on a state-of-the-art CNN. We conclude that the introduced techniques offer promising improvements over the baseline in almost all cases, and that the selection of the technique may vary depending on the amount of data available and the level of imbalance.
\end{abstract}


%

\end{frontmatter}

\section{Introduction}

Deep learning algorithms exhibit remarkable capabilities in computer-aided detection and diagnosis (CAD) across diverse applications~\cite{CHEN2022102444}, including disease classification~\cite{IMPORTANTEshorfuzzaman2021metacovid, Wang2017chestX, narin2021automatic}, segmentation~\cite{IMPORTANTEfan2020inf, Havaei2017tumorSegment}, or medical object detection such as pulmonary nodules~\cite{mei2021sanet} or lymphocytes~\cite{swiderska2019learning}, among others. In particular, the emergence of annotated X-ray imaging datasets~\cite{padchest, bimcv-covid, chestnihcc} has made the research of many applications based on deep neural networks possible, greatly benefiting pathology diagnosis and prognosis.

Nevertheless, the performance of models trained on medical images highly depends on several factors that can notably worsen the results. Key challenges include the scarcity of annotated data and the substantial cost associated with expert labeling~\cite{varoquaux2022machine}. Compared to regular datasets in computer vision, a medical image dataset usually contains relatively few images, and in some cases, only a small percentage of them are annotated by experts~\cite{CHEN2022102444}. In addition, there is commonly a considerable imbalance between negative (healthy) and positive (pathological) samples. Moreover, generated models strongly rely on the specific domain of data for which they were trained. All these challenges collectively hinder the development of effective, robust, and generalizable methods for processing medical images~\cite{Razzak2018}, and only a few approaches based on deep learning techniques are eventually certified for clinical usage~\cite{sendak2020path}.

A standard solution to deal with the scarcity of annotated medical imaging data due to its associated high costs is data augmentation~\cite{shorten2019survey, Garay2019vae}. This technique generates synthetic samples from existing images, expanding the training dataset. However, the distinctive characteristics of medical images, such as their high dimensionality, intricate structures, and substantial inter- and intra-class variability, present challenges when applying traditional data augmentation techniques~\cite{GARCEA2023106391}. Therefore, designing effective augmentation strategies for medical imaging often requires domain expertise involving radiologists or medical professionals who can provide guidance and validation.

Another widely adopted solution for addressing the limited availability of annotated data is using transfer learning~\cite{weiss2016survey}. This technique involves leveraging knowledge acquired from a domain with sufficient labeled data and applying it to another domain by fine-tuning the model. In this process, the weights of a pre-trained model are used as an initialization for a new model. Transfer learning has gained significant interest for medical imaging~\cite{kim2022transfer, Shin2016tl, Ghafoorian2017}. Notably, it helps reduce the amount of labeled data required for training, accelerates convergence, and yields models with better generalization capabilities. These generalized models can be effectively transferred to other domains, enabling inter-domain use.

The issue of highly imbalanced data is another common challenge in medical imaging, where the number of positive samples is often significantly lower than that of negative ones~\cite{ramyachitra2014imbalanced}. Machine learning models trained on imbalanced data tend to exhibit bias towards the majority class, not paying attention to the samples from the minority class~\cite{He2009imbalanced}. Consequently, this leads to suboptimal performance for the underrepresented samples, which can have severe consequences in detecting specific pathologies and could represent a risk for the patients in critical scenarios. 

A small dataset becomes even more prone to overfitting, making the model lose generalization capabilities when the training data is not large enough. Few-shot learning (FSL) algorithms address this issue. These methods can be categorized~\cite{weng2020addressing} into metric-based, optimization-based, and transfer learning-based approaches. Metric-based FSL learns a representation by comparing training examples through Siamese networks~\cite{koch2015siamese}, matching networks~\cite{vinyals2016matching}, prototypical networks~\cite{snell2017prototypical}, or relation networks~\cite{sung2018learning}. Optimization-based FSL~\cite{finn2017model} can learn the parameters of any standard model via meta-learning in such a way as to prepare that model for fast adaptation. These techniques include Model-Agnostic Meta-Learning (MAML)~\cite{finn2017model}, LSTM-based meta-learner models~\cite{ravi2016optimization}, and Proto-MAML~\cite{triantafillou2019meta}. Finally, transfer learning-based approaches include fine-tuning~\cite{ChenLKWH19} and linear models learned on top of a pre-trained embedding~\cite{tian2020rethinking}, such as $k$-Nearest Neighbor ($k$NN)~\cite{knn}, Support Vector Machine (SVM)~\cite{svm}, or Random Forest (RF)~\cite{rf}.


Although FSL has been studied extensively, only a few of these techniques~\cite{nayem2023few} have been investigated for medical imaging. In~\cite{fewshotm2}, a MAML algorithm is adopted for a few-shot problem with medical images, and the Dice loss function is used to mitigate class imbalance. Different FSL methods are compared in~\cite{weng2020addressing} for the skin condition recognition problem in which class imbalance exists, showing that when combined with conventional imbalance techniques, they lead to better performance, especially for the rare classes.

The main objective of this work is to investigate the accuracy of learning-based models in the medical imaging domain, focusing on their behavior in few-shot and imbalanced scenarios. In~\cite{ibpria}, we studied the effect of different techniques to deal with imbalanced data but for scenarios with sufficient samples. The evaluation was performed on different chest X-ray datasets labeled with COVID-19 positive and negative diagnoses. Here, we extend this previous work by proposing and evaluating similar techniques but adapted to the few-shot learning paradigm with imbalanced data. In particular, we use a metric-based FSL method based on Siamese networks~\cite{Valero-Mas2023} in which a series of proposals are integrated to mitigate the effects of few and imbalanced data, including different initialization methods, transfer learning, data augmentation, four proposals adapted to Siamese neural networks to deal with imbalanced data, and four alternative classifiers to carry out the final prediction.

To carry out the evaluation, four publicly available chest X-ray image datasets~\cite{padchest, bimcv-covid, chestnihcc, githubcovid} are considered. Three corpus pairs are created from these, each containing positive and negative samples of COVID-19 patients. The performance of these techniques is evaluated in both intra-domain (within the same domain) and inter-domain (across different domains) use cases, and for four levels of data imbalance. The results of the different experiments carried out show that the low number of parameters due to the shared weights of both Siamese networks, along with the included proposals, improve the results, reduce the tendency to overfit and the amount of data required for training. 

The remainder of the paper is organized as follows:
Section~\ref{sec:methodology} outlines the proposed approach to address the challenges discussed earlier; 
Section~\ref{sec:setup} presents the experimental setting used to evaluate the approach, including details about the datasets used for experimentation;
Section~\ref{sec:results} presents and analyzes the evaluation results obtained from applying the proposed techniques; and 
Section~\ref{sec:conclusions} finally concludes the paper by summarizing the key findings and contributions of the study. Additionally, it outlines potential directions for future research in the medical imaging domain and the challenges that remain to be addressed.

\section{Methodology}
\label{sec:methodology}

This section describes the methodological proposal to address the challenges that learning-based methods commonly face when dealing with medical image datasets, which, as mentioned, are mainly data scarcity and intrinsic imbalance according to the data distribution.

\begin{figure}[ht]
    \centering
    \includegraphics[width=\textwidth]{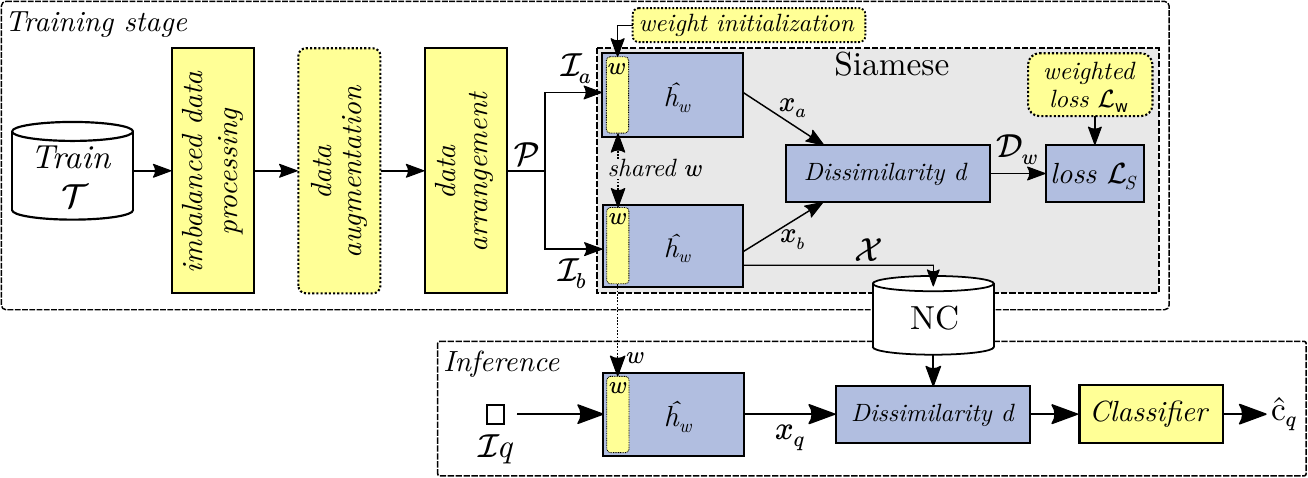}
    \caption{Diagram with the pipeline of the process. The proposed techniques to be studied are highlighted in yellow.}
    \label{fig:method}
\end{figure}

Figure~\ref{fig:method} illustrates the pipeline steps followed during the training and inference stages. Formally, let $\mathcal{T} = \left\{\left(I_{i}, c_{i}\right): I_{i}\in\mathcal{I}, c_{i}\in\mathcal{C}\right\}_{i=1}^{\left|\mathcal{I}\right|}$ represent a set of labeled images where $\mathcal{I}$ denotes the input data space and $\mathcal{C}$ the set of possible categories. Let also $\zeta: \mathcal{I}\rightarrow \mathcal{C}$ be the function that relates the input image $I_{i}$ with its associated class $c_{i}$, i.e., $\zeta\left(I_{i}\right) = c_{i}$.

During the training phase, the aim is to learn an approximation of $\zeta$, denoted by $\hat{h}_w$, which is implemented through a learning-based network parameterized with a set of weights $w$. To learn $\hat{h}_w$, the training set $\mathcal{T}$ is used to minimize the network error according to a given loss function $\mathcal{L}$. This work analyzes the improvement brought to this learning process by different techniques that address the challenges posed. 

In the proposed pipeline, input data is first processed to balance the sampling carried out and adjust the data distribution of $\mathcal{T}$. A data augmentation process is also considered to generate more training samples artificially. This preprocessed data is then used to learn the function $\hat{h}_w$, for which a Siamese architecture is considered, as it is specially devised for few-shot scenarios. Different initialization techniques are also studied in this step, including transfer learning. Besides, a weighted loss function $\mathcal{L}_w$ is introduced to address the imbalance and improve the model training further.

Once the training is completed, the inference stage is carried out. Given a set of query data $\mathcal{Q} = \left\{\left(I_{q}\right)\right\} \subset\mathcal{I}\times\mathcal{C}$, inference is performed by considering the estimated function $\hat{h}_w$ to calculate the final prediction $\hat{c}_q$, i.e., $\hat{h}_w\left(I_{q}\right) = \hat{c}_q$. For this, a new model $\hat{h}_w$ is generated from the weights $w$ of one of the parallel networks of the Siamese architecture. The query sample $I_q$ is then processed by the network to extract its embedding representation, which is compared with the embeddings (also called Neural Codes or NC) obtained for the training set $\mathcal{T}$ to compute the final prediction.

The following sections provide a detailed explanation of each step of this process, starting with the definition of the Siamese architecture.

\subsection{Siamese architecture}

The Siamese architecture consists of two identical parallel networks with shared weights, which process two input images to determine whether they are equal. This configuration is especially suitable for few-shot learning scenarios due to two main reasons. On the one hand, it simplifies the task as it only aims to determine the similarity of the images and not the class. On the other hand, the pair-wise arrangement of the set $\mathcal{T}$ increases the number of samples used to train the model (in practice, $M = \binom{|\mathcal{T}|}{2}$ possible pairs may be generated). Therefore, this arrangement results in greater variability of input data, which favors the convergence of the neural scheme.

Let $\mathcal{P} = \left\{\left(\left\{I_{a}, I_{b}\right\}, y_i\right): \left\{I_{a}, I_{b}\right\}\in\mathcal{I}, y_{i}\in\mathcal{Y}\right\}_{i=1}^{M}$ represent the set with all possible pairs of images $\left\{I_{a}, I_{b}\right\}$ drawn from the defined input space $\mathcal{I}$ and $y_{i}\in\mathcal{Y}$ be a binary indicator depicting whether the input pair is similar or different. The Siamese architecture initially maps the input pair $\left\{I_{a}, I_{b}\right\}$ using the networks $h_w$ to a new $N$-dimensional space $\mathcal{X} \in \mathbb{R}^{N}$, obtaining the feature vectors $\mathbf{x}_{a}$ and $\mathbf{x}_{b}$, respectively. In this new space, given a dissimilarity metric $d : \mathcal{X} \times \mathcal{X} \rightarrow \mathbb{R}^{+}_{0}$, a similitude score $D_{w}$ between $\mathbf{x}_{a}$ and $\mathbf{x}_{b}$ is calculated. This value is meant to be zero when the images are equal and move away proportionally according to the degree of dissimilarity. Note that $D_{w}$ should be thresholded (either heuristically or with a learning-based method) to establish whether the inputs are similar. The block labeled ``Siamese'' in Figure~\ref{fig:method} shows a graphical scheme of this architecture.

The Siamese networks are trained using the so-called \textit{contrastive loss} which, for a single pair of data $\left(I_{a}, I_{b}\right)$, is defined as:

\begin{equation}
\mathcal{L}\left(w,\left(y,I_{a},I_{b}\right)\right) 
= 
\left(1-y\right) \cdot D^{2}_{w} + y \cdot \max\left(0, m - D_{w} \right)^2 
\label{eq:contrastive_loss}
\end{equation}

\noindent 
where $D_{w}$ stands for the dissimilarity value between input elements, i.e., $D_{w} = d\left(x_{a},x_{b}\right)$, $y$ for the binary class-matching indicator, and $m$ represents a separation margin following the proposal by Hadsell et al.~\cite{hadsell2006dimensionality} to define a \textit{hinge} or \textit{maximum margin} loss.

From this, the total loss $\mathcal{L}_{S}$ can be calculated as the sum of the partial losses for each pair in $\mathcal{P}$, i.e., $\mathcal{L}_{S} = \sum_{i=1}^{|\mathcal{P}|}\mathcal{L}\left(w,\left(y, I_{a}, I_{b}\right)^{(i)}\right)$.

In this context, this work studies the performance of this scheme in imbalance few-shot scenarios and the improvement that different additional mechanisms bring to this process, such as initialization techniques, transfer learning, data augmentation, and proposals to balance the data distribution, as introduced in the following sections.

\subsection{Siamese Initialization}

In a few-shot learning scenario, the initialization of the neural network weights plays a crucial role since it can influence both the final result and the number of samples needed for training~\cite{Raghu2019tl}. To assess its effect on the task at hand, three initialization strategies are studied:

\begin{itemize}
    \item Training from scratch: The network is initialized with random weights, leading to a learning process that begins from scratch. This approach typically requires a larger set of labeled data for the model training to converge.

    \item Initializing the network with ImageNet pre-trained weights: Although it is a very different domain, leveraging knowledge from this large-scale dataset reduces the training time and data requirements, potentially accelerating the learning process and improving the results obtained. 

    \item Transfer learning: This approach initializes the network using the weights obtained with a similar X-ray dataset for which there is a larger availability of labeled data and then applies a fine-tuning process to the target distribution. In this way, training starts from a good initialization and can benefit from the knowledge extracted from a closer domain while adapting to the particularities of the new data. Note that, in this case, due to the larger quantity of data, the initial training may be carried out on the $h_w$ backbone used in the Siamese (without pairwise training) and then construct the Siamese architecture from this.
\end{itemize}

\subsection{Data augmentation} 
\label{sec:dataaug}

Data augmentation has become a \textit{de facto} standard in training learning-based methods due to its good results. This technique increases the size and diversity of a training dataset by applying transformations to the existing samples, which may include rotations, skew, scaling, cropping, flipping, and contrast or color adjustments, among others. The introduced variability improves the trained models' robustness and generalizability and reduces overfitting, making it a valuable tool for small training sets.

However, the effectiveness of each transformation largely depends on the specific task to be solved. In the context of medical imaging, its unique properties require a more cautious approach when applying data augmentation~\cite{shorten2019survey, Chlap2021daug}. Some inappropriate transformations can hide or alter certain findings that could be key to diagnosing a pathology (for example, a flip operation would change the heart's position). Consequently, we have considered a limited set of transformations that do not alter the shape or invert the position of elements in the image. Specifically, the effect of the following set of transformations is studied as the value of the $\alpha$ parameter increases:

\begin{itemize}
    \item Horizontal and vertical shifts (in the range of $[-\alpha, \alpha]$\% of the image size).
    \item Scaling (in the range of $[-\alpha, \alpha]$\% of the original image size).
    \item Rotations (in the range of $[-\alpha^\circ, \alpha^\circ]$).
\end{itemize}

\subsection{Imbalanced data}
\label{sec:method:imbalanced}

While previous sections have focused on solutions for small training sets, this section describes the techniques aimed at dealing with data imbalance. For this, four proposals are assessed: balancing the sample distribution, weighting the loss function, combining balancing with the loss, and modifying the ratio of positive and negative pairs. Note that when we talk about positive and negative pairs in the Siamese network, we mean pairs of images that belong to the same class and pairs of images of different classes, respectively, regardless of whether they represent sick or healthy cases.

As previously indicated, the total number of training pairs is calculated as $M = \binom{|\mathcal{T}|}{2}$, which may be decomposed into $M = \binom{|\mathcal{T}_P|}{2} + \binom{|\mathcal{T}_N|}{2}$, where $\mathcal{T}_P$ and $\mathcal{T}_N$ represent the total number of samples that could form positive and negative pairs, respectively. From this, we can calculate the imbalance ratio as $r=\nicefrac{|\mathcal{T}_P|}{|\mathcal{T}_N|}$ and increase the sampling of the minority class until $r=1$. This \textit{balanced sampling} proposal is equivalent to the \textit{Oversampling} technique studied in our previous work~\cite{ibpria} since it consists of duplicating the samples of the minority class but in a way adapted to Siamese networks. In this case, \textit{Undersampling} is not considered, since it has been proven to yield poor results, which would be even worse in this scenario with few data. 

A second proposal to deal with imbalanced distributions is to \textit{weight the loss function} during the training stage. Specifically, this technique increases the value of the error committed for the minority classes to balance their contribution to the overall error. This forces the training process to treat all classes equally and prevents creating a bias towards the majority class. As far as we know, there are no proposals to weight the contrastive loss used by Siamese networks. For this reason, we propose to modify Equation~\ref{eq:contrastive_loss} by introducing the following weighting factor:

\begin{equation}
\mathcal{L}_w = \frac{\lambda_{\zeta(I_a)} + \lambda_{\zeta(I_b)}}{2} \left( \left(1-y\right) \cdot (D_{w}(I_a, I_b))^{2} + y \cdot \max\left(0, m - D_{w}(I_a, I_b) \right)^2 \right)
\label{eq:weighted_contrastive_loss}
\end{equation}

\noindent
where the parameters $\lambda_{c_i}$ represent the factors used to weight the classes $c_i$ of each sample $I_a$ and $I_b$, respectively, recovered as $c_a = \zeta(I_a)$ and $c_b = \zeta(I_b)$. $\lambda_{c_i}$ is calculated as the quotient of the total training samples $|\mathcal{T}|$ divided by the number of classes $|\mathcal{C}|$ multiplied by the number of samples of the class $c_i$, i.e. $|\mathcal{T}|_{c_i}$. This weighting factor can be expressed as:

\begin{equation}
\lambda_{c_i} = \frac{|\mathcal{T}|}{|\mathcal{C}| \cdot |\mathcal{T}|_{c_i}}
\end{equation}

As a third proposal, we will study the effect of applying the balanced sampling and the weighted loss function in a combined manner.

Finally, it is also proposed to modify the \textit{balance of pairing} of positive and negative examples used during network training. That is, instead of generating a set $\mathcal{P}$ with the same number of positive and negative pairs, it is proposed to change this proportion so that the network, for example, sees many more negative pairs than positive ones (or vice versa). This technique also modifies the distribution of the data, as it requires drawing a sample from each class to create negative pairs, and consequently, the instances from the minority class will be repeated.

\subsection{Inference stage}
\label{sec:method:inference}

The Siamese architecture is designed to determine a similarity score that correlates the embedded representations of input elements rather than directly retrieving class labels for classification tasks. Therefore, the following procedure is usually considered to adapt Siamese schemes for classification purposes: given a query sample denoted as $I_{q}$, the distances between this item and the entire training set $\mathcal{T}$ are computed in the embedded representation space $\mathcal{X}$. The query $I_{q}$ is eventually assigned with the label $\hat{c}_{q}$, which corresponds to the label of the element that exhibits the minimum distance value. This process can be expressed as follows:

\begin{equation}
\hat{c}_{q} = \zeta\left(\argmin_{\forall I_{i}\in\mathcal{T}} d\left(h_w(I_{q}), h_w(I_{i})\right)\right)
\end{equation}

In addition to this approach (which we will refer to as Histogram), it is proposed to study the improvement provided by the use of a model learned using the embeddings generated by the Siamese network, a technique that could be considered a transfer-learning approach according to the literature~\cite{tian2020rethinking}. Specifically, the trained $h_w$ network is used to transform the inputs to the embedded representation space $\mathcal{X}$, on which three alternative methods are applied to calculate the final correlation:

\begin{itemize}
    \item $k$-Nearest Neighbor ($k$NN)~\cite{knn}: This algorithm categorizes the given query $I_{q}$ by identifying the prevailing class among the $k$ nearest elements to it. For this, a dissimilarity metric is used to compare the embedding of the query with those of the training set (NC in Figure~\ref{fig:method}).
    
    \item Support Vector Machine (SVM)~\cite{svm}: This approach transforms the original data into a higher-dimensional space using a specified kernel function. Subsequently, it learns a hyperplane to distinguish between the classes. 
        
    \item Random Forest (RF)~\cite{rf}: This method constructs an ensemble classifier from individual decision trees, each trained on random data subsets. The final output amalgamates the decisions from each tree in order to calculate the class of the input query.
\end{itemize}

\section{Experimental setup}
\label{sec:setup}

This section details the experimental setup, including the selection of datasets, the network architecture and the parameters chosen, the training process details, and the evaluation metrics employed.

\subsection{Datasets}
\label{sec:datasets}

\begin{figure}[!ht]
    \centering
    \begin{subfigure}[b]{.24\textwidth}
        \centering
        \includegraphics[width=\textwidth]{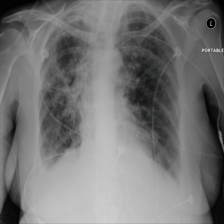}
        \caption{ChestX-ray}
        \label{fig:sub:chestx}
    \end{subfigure}
    \begin{subfigure}[b]{.24\textwidth}
        \centering
        \includegraphics[width=\textwidth]{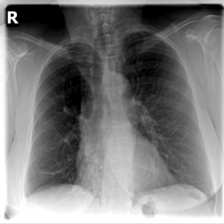}
        \caption{PadChest}
        \label{fig:sub:padchest}
    \end{subfigure}
    \begin{subfigure}[b]{.24\textwidth}
        \centering
        \includegraphics[width=\textwidth]{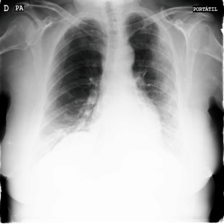}
        \caption{BIMCV-COVID}
        \label{fig:sub:BIMCV}
    \end{subfigure}
    \begin{subfigure}[b]{.24\textwidth}
        \centering
        \includegraphics[width=\textwidth]{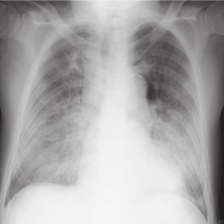}
        \caption{Github-COVID}
        \label{fig:sub:github}
    \end{subfigure}
    \caption{Illustrative samples from the evaluated datasets.}
    \label{fig:datasets}
\end{figure}

The methodology was assessed using four distinct datasets\footnote{All datasets are publicly accessible: ChestX-ray can be found at \url{https://nihcc.app.box.com/v/ChestXray-NIHCC}, GitHub-COVID at \url{https://github.com/ieee8023/covid-chestxray-dataset}, PadChest is available at \url{https://bimcv.cipf.es/bimcv-projects/padchest}, and BIMCV-COVID repositories can be accessed through \url{https://bimcv.cipf.es/bimcv-projects/bimcv-covid19}.}. An overview of these datasets is presented in Table~\ref{tab:datasets1}, indicating the types of samples they contain (negative ($-$) or positive ($+$) COVID-19 samples), along with the original sizes of the training and evaluation sets. Example images from these datasets are shown in Figure~\ref{fig:datasets}.

\begin{table}[!ht]
    \caption{The initial configuration of the datasets under consideration is as follows, showing the type of samples (positive $+$ and negative $-$ COVID-19 patients), the number of samples per class, and their total ($\sum$). Additionally, the size of the training and test sets is provided, along with the percentage of each set compared to the total size.}
    \label{tab:datasets1}
    \setlength{\tabcolsep}{6pt}
    \centering
    \begin{tabular}{lcrlrlc}
    \toprule
    \textbf{Dataset}              
        & \textbf{Classes}    
        & \multicolumn{2}{c}{\textbf{Train size}} 
        & \multicolumn{2}{c}{\textbf{Test size}}
        & \textbf{Total}  \\ 
    \midrule
    ChestX-ray~\cite{chestnihcc}  & $-$       & 86\,524 & (77\%)       & 25\,596 & (23\%)      & 112\,120 \\ 
    \hline
    PadChest~\cite{padchest}     & $-$       & 91\,508 & (95\%)       & 4\,762  & (5\%)       & 96\,270  \\ 
    \hline
    \multirow{3}{*}{BIMCV-COVID~\cite{bimcv-covid}}    
                  & $-$       & 3\,014  &              & 159     &             & 3\,173   \\
                  & $+$       & 1\,610  &              & 82      &             & 1\,692   \\
                  & $\sum$    & 4\,624  & (95\%)       & 241     & (5\%)       & 4\,865   \\ 
    \hline
    \multirow{3}{*}{Github-COVID~\cite{githubcovid}} 
                  & $-$       & 81      &              & 29      &             & 110    \\
                  & $+$       & 283     &              & 11      &             & 294    \\
                  & $\sum$    & 364     & (90\%)       & 40      & (10\%)      & 404    \\ 
    \bottomrule
    \end{tabular}
\end{table}

As it can be seen in Table~\ref{tab:datasets1}, two of the datasets exclusively contain negative samples of COVID-19 patients, while the other two, although comprising both classes, exhibit class imbalance. To evaluate the proposed methodology, three combinations were made from these data, creating new datasets with both positive and negative samples, as presented in Table~\ref{tab:datasets2}. This table introduces an acronym for each combination (to be used in the experimentation section) and specifies the number of positive and negative samples in each newly generated set. As in the previous work~\cite{ibpria}, the number of samples added from the original datasets was limited to 10,000 to ease the experiments. Additionally, the ``mean imbalance ratio'' (MeanIR) index is provided to indicate the imbalance level of the corpus~\cite{ValeroMas2023mpg}. The MeanIR value ranges from $\left[1, \infty\right)$ and denotes a higher imbalance as the value increases.

\begin{table}[!ht]
    \caption{Description of the new combined datasets derived from Table~\ref{tab:datasets1}. They include the acronym, partition sizes, the count of positive ($+$) and negative ($-$) COVID-19 samples, and their respective percentages. The MeanIR, an indicator of dataset balance, is also provided.}
    \label{tab:datasets2}
        \footnotesize
        \centering
        \setlength{\tabcolsep}{2.5pt}
        \resizebox{\textwidth}{!}{\begin{tabular}{lcrlcrlcrlcc}
            \toprule
            \textbf{Acronym}                     
                & \textbf{Combined data}
                & \multicolumn{2}{c}{\textbf{Train size}} 
                &  & \multicolumn{2}{c}{\textbf{Test size}} 
                &  & \multicolumn{2}{c}{\textbf{Total}} 
                &  & \textbf{MeanIR}  \\ 
            \midrule
            \multirow{3}{*}{ChestX-Git} 
                & \multirow{3}{*}{\begin{tabular}[c]{@{}c@{}}ChestX-ray\\ $\cup$ \\Github-COVID\end{tabular}} 
                    & $-$:              & 10\,081             
                        &  & $-$:       & 10\,029             
                        &  & $-$:       & 20110 (99\%)      
                        &  &            \multirow{3}{*}{34.7} \\
                    &  & $+$:           & 283               
                        &  & $+$:       & 11               
                        &  & $+$:       & 294 (1\%)         
                        &  &    \\
                    &  & $\sum$:        & 10\,364 (51\%)      
                        &  & $\sum$:    & 10\,040 (49\%)      
                        &  & $\sum$:    & 20\,404             
                        &  &   \\ 
            \midrule
            \multirow{3}{*}{Pad-BIM}    
                & \multirow{3}{*}{\begin{tabular}[c]{@{}c@{}}PadChest\\ $\cup$ \\BIMCV-COVID+\end{tabular}}              
                    & $-$:       & 10\,000             
                        &  & $-$:       & 4\,762              
                        &  & $-$:       & 14\,762 (90\%)      
                        &  &            \multirow{3}{*}{4.9} \\
                    &  & $+$:           & 1\,610              
                        &  & $+$:       & 82  
                        &  & $+$:       & 1\,692 (10\%)     
                        &  &    \\
                    &  & $\sum$:        & 11\,610 (71\%)      
                        &  & $\sum$:    & 4\,844 (29\%)       
                        &  & $\sum$:    & 16\,454           
                        &  &    \\ 
            \midrule
            \multirow{3}{*}{BIMCV-COVID}      
                & \multirow{3}{*}{\begin{tabular}[c]{@{}c@{}}BIMCV-COVID-\\ $\cup$ \\BIMCV-COVID+\end{tabular}}      
                    & $-$:              & 3\,014              
                        &  & $-$:       & 159               
                        &  & $-$:       & 3\,173 (65\%)       
                        &  &            \multirow{3}{*}{1.4} \\
                    &  & $+$:           & 1\,610              
                        &  & $+$:       & 82  
                        &  & $+$:       & 1\,692 (35\%)       
                        &  &  \\
                    &  & $\sum$:        & 4\,624 (95\%)       
                        &  & $\sum$:    & 241 (5\%)         
                        &  & $\sum$:    & 4\,865              
                        &  &  \\ 
            \bottomrule 
        \end{tabular}}%
\end{table}

Note that since the size of the training partitions in these corpora does not meet the requirements of a few-shot learning scenario, we artificially reduce their size while leaving the test sets unaltered. Specifically, for the experimentation, 10-fold cross-validation was carried out, selecting for each fold 100 random samples without repetition from the majority class (healthy patients) and $n$ random samples from the minority class (COVID-19+ patients). For the value of $n$, four possible imbalance scenarios were considered: \textit{H}igh imbalance with $n=1$, \textit{M}edium imbalance with $n=10$, \textit{L}ow imbalance with $n=50$, and \textit{N}o imbalance with $n=100$. In addition, the effect of the proposed techniques is also studied when the number of samples is increased to 200 and 300 while maintaining the level of imbalance. Note that, in all cases, the evaluation was carried out with the complete test set as indicated in Table~\ref{tab:datasets2}.

\subsection{Network architecture}

The proposed methodology was assessed using ResNet-50 v2~\cite{He2016ResNet50v2} as the backbone for the $h_w$ Siamese parallel networks. This is a standard architecture for image classification known for its state-of-the-art results in various benchmarks and applications~\cite{Gallego2021idann}. This updated version of ResNet-50 incorporates identity shortcuts and pre-activation units, enhancing performance and reducing overfitting.

Regarding the rest of the configuration details of the Siamese architecture, the Euclidean distance was considered as dissimilarity function $d$ (i.e., $D_w = \sqrt{(h_w(I_a)-h_w(I_b))^2}$) and the $\ell_2$ normalization~\cite{ZhengZWWT16} for the regularization of the embedded representations. 

For the margin parameter $m$ of the loss function (see Equation~\ref{eq:contrastive_loss}), initial experimentation was carried out considering values in the range $m \in [0, 8]$, obtaining low results for the extremes of this range. The value of $m=1$ was eventually selected for the rest of the experimentation, as it reported the best results overall.

Throughout all the experiments, the Siamese networks were trained for 200 epochs with a batch size of 32 images. Stochastic Gradient Descent~\cite{Mitchell1997MachineLearning} was employed for parameters optimization with a Nesterov momentum of 0.9, a learning rate of $10^{-2}$, and a decay factor of $10^{-6}$. The images were scaled to 224$\times$224 pixels, and their values were normalized within the range $[0, 1]$ to aid model convergence.

\subsection{Metrics}

For the quantitative evaluation, we used the F-measure ($\mbox{F}{1}$) as the figure of merit to mitigate potential biases caused by significant label imbalances in the considered datasets. In a binary classification scenario, $\mbox{F}{1}$ is calculated as the harmonic mean of Precision ($P$) and Recall ($R$). The definitions of these metrics are as follows:

\begin{align}
    \mbox{P} &= \frac{\mbox{TP}}{\mbox{TP} + \mbox{FP}}\label{eq:P}\\
    \mbox{R} &= \frac{\mbox{TP}}{\mbox{TP} + \mbox{FN}}\label{eq:R}\\
    \mbox{F}_{1} &= \frac{2 \cdot \mbox{P} \cdot \mbox{R}}{\mbox{P} + \mbox{R}} = \frac{2\cdot\mbox{TP}}{2\cdot\mbox{TP} + \mbox{FP} + \mbox{FN}}\label{eq:F1}
\end{align}

\noindent 
where TP, FP, and FN denote the number of true positives, false positives, and false negatives, respectively.

The evaluation involved multi-class experiments, so the results are reported in terms of macro-$\mbox{F}_{1}$ for a comprehensive assessment. Macro-$\mbox{F}_{1}$ is computed as the average of the $\mbox{F}_{1}$ scores obtained for each class.

\section{Results}
\label{sec:results}

In this section, the proposed methodology is evaluated using the datasets, network configuration, and metrics described previously. To provide a comprehensive assessment, the results of each technique presented before, applied on the network of Figure~\ref{fig:method}, are analyzed individually. The section starts with the effects of the initialization process, then delves into data augmentation analysis, contrasts techniques for data imbalance, compares inference classifiers, and examines the influence of training set size. Finally, the section includes a discussion with concluding remarks, comparing the few-shot learning scenario with the results from the prior study that explored techniques without labeled data constraints.

In all cases, results are analyzed at intra- and inter-domain levels, as well as for four imbalanced data distributions. These distributions are referred to with the initials \textit{H}, \textit{M}, \textit{L}, and \textit{N}, being \textit{H} $\rightarrow$ \textit{H}igh imbalance (100/1), \textit{M} $\rightarrow$ \textit{M}edium (100/10), \textit{L }$\rightarrow$ \textit{L}ow (100/50), and \textit{N} $\rightarrow$ \textit{N}one (100/100).

\subsection{Initialization} 
\label{sec:initialization}

As a recap of the pipeline presented in the methodology, one way to cope with small sets of data is the use of a good initialization of the network weights before starting the training process. In this section, we will focus on studying the effects of different initialization techniques. First, a baseline result is obtained by training the Siamese ResNet-50 v2 backbones from scratch, i.e., using random values as initialization parameters. It is compared to a pre-initialized model whose weights are obtained from a generic dataset, in this case, ImageNet~\cite{Deng2009imagenet}, that, afterward, is fine-tuned with our datasets. For the sake of simplicity, we will refer to them as scratch and pre-initialized models.

Table~\ref{tab:result:ini:1} shows the macro-$\mbox{F}_{1}$ results of both approaches, scratch and pre-initialized with weight initialization, for the four levels of data imbalance considered. This table also includes detailed results for each possible training-to-evaluation dataset combination. The ``From'' column indicates the training source, whereas ``To'' refers to the evaluation set. Hence, we evaluate cases within the same domain (intra-domain), which are underlined, and inter-domain cases where the model is assessed on domains different from its training source. The best result per experiment and imbalance level is marked in bold, i.e., the best figure obtained according to the initialization method, either from scratch or pre-initialized. For instance, the value 45.1 appears in bold in the first column (corresponding to the BIMCV-COVID test set) because the training from scratch approach is better than the weight initialization (which obtains a 44.7 in this case). On the contrary, the pre-initialized model achieves higher performance in the high imbalance cases for the Chest-Git (with 42.2) and Pad-BIM (with 46.0) test sets. 

From a global perspective, the results show that, in most cases, the performance of the pre-initialized network achieves better results, especially for the cases with \textit{H}igh, \textit{M}edium, and \textit{L}ow imbalance. Regarding the \textit{N}one imbalanced experiments, the results obtained are quite similar for both initialization approaches. This makes it clear that the architecture presented can learn efficiently regardless of initialization, even for this low-data scenario. The high variability generated by possible combinations of training pairs makes it not so dependent on initialization. However, in the case of \textit{H}igh imbalance, this architecture appears to struggle with convergence during training, as a single example from the minority class may be insufficient. These results improve progressively as the level of imbalance decreases. It is also noteworthy that the average intra-domain results are promising starting from a \textit{M}edium imbalance, especially considering that it is a few-shot scenario.

\begin{table}[htb]
    \centering
    \caption{$\mbox{F}_{1}$ results achieved by training the model from scratch and initializing with ImageNet weights. In each scenario, the intra-domain cases are underlined for clarity. Each case is analyzed considering four levels of imbalance: \textit{H}igh, \textit{M}edium, \textit{L}ow, and \textit{N}one.}
    \label{tab:result:ini:1}
    \setlength{\tabcolsep}{4pt}
    \footnotesize
        \resizebox{\textwidth}{!}{\begin{tabular}{lllcccclcccc} 
            \toprule
             &  &  & \multicolumn{4}{c}{From scratch} 
                &  & \multicolumn{4}{c}{Weight initialization} \\ 
            \cmidrule{4-7}\cmidrule{9-12}
            \textbf{From} & \textbf{To} 
                &  & \textit{H} & \textit{M} & \textit{L} & \textit{N} 
                 &  & \textit{H} & \textit{M} & \textit{L} & \textit{N} \\ 
            \cmidrule{1-2}\cmidrule{4-7}\cmidrule{9-12}
            
                \multirow{3}{*}{Chest-Git} 
                    & \uline{Chest-Git} 
                        &  & 39.2           & 52.4          & 68.0          & \textbf{74.8} 
                        &  & \textbf{42.2}  & \textbf{61.7} & \textbf{72.2} & 73.2 \\
                    & Pad-BIM 
                        &  & 44.9           & 54.8          & \textbf{61.2} & \textbf{61.3} 
                        &  & \textbf{46.0}  & \textbf{60.4} & 57.4          & 56.9 \\                    
                    & BIMCV-COVID 
                        &  & \textbf{45.1}  & 45.5          & \textbf{47.1} & \textbf{47.3} 
                        &  & 44.7           & \textbf{49.7} & 46.9          & 46.3 \\ 
            
            \cmidrule{1-2}\cmidrule{4-7}\cmidrule{9-12}

                \multirow{3}{*}{Pad-BIM} 
                    & Chest-Git 
                        &  & 42.4           & \textbf{51.0} & 44.7          & \textbf{45.8} 
                        &  & \textbf{43.6}  & 49.7          & \textbf{47.6} & 43.4 \\
                    & \uline{Pad-BIM} 
                        &  & 50.9           & 54.9          & 70.6          & 81.8 
                        &  & \textbf{55.4}  & \textbf{63.2} & \textbf{79.2} & \textbf{82.8} \\                    
                    & BIMCV-COVID 
                        &  & \textbf{47.8}  & 46.1          & 51.5          & \textbf{53.0} 
                        &  & 46.3           & \textbf{51.4} & \textbf{54.4} & 50.8 \\ 
                    
            \cmidrule{1-2}\cmidrule{4-7}\cmidrule{9-12}
            
                \multirow{3}{*}{BIMCV-COVID}                     
                    & Chest-Git 
                        &  & 42.3           & \textbf{53.7} & 44.7          & \textbf{52.7} 
                        &  & \textbf{44.0}  & 52.8          & \textbf{54.2} & 48.8 \\
                    & Pad-BIM 
                        &  & \textbf{51.2}  & 52.2          & 46.3          & 54.0 
                        &  & 48.1           & \textbf{55.3} & \textbf{58.7} & \textbf{55.9} \\
                    & \uline{BIMCV-COVID} 
                        &  & \textbf{48.4}  & \textbf{47.8} & 49.1          & \textbf{55.1}
                        &  & 47.3           & 46.4          & \textbf{51.9} & 50.9 \\ 
                
            \cmidrule[\heavyrulewidth]{1-2}\cmidrule[\heavyrulewidth]{4-7}\cmidrule[\heavyrulewidth]{9-12}
            \multicolumn{2}{r}{\textbf{Inter-Domain Avg.}} 
                &  & \textbf{45.6}  & 50.5          & 49.2          & \textbf{52.3} 
                &  & 45.4           & \textbf{53.2} & \textbf{53.2} & 50.3 \\
            \multicolumn{2}{r}{\textbf{Intra-Domain Avg.}} 
                &  & 46.2           & 51.7          & 62.6          & \textbf{70.5} 
                &  & \textbf{48.3}  & \textbf{57.1} & \textbf{67.8} & 68.9 \\
            \multicolumn{2}{r}{\textbf{Global Avg.}} 
                &  & 45.8           & 50.9          & 53.7          & \textbf{58.4} 
                &  & \textbf{46.4}  & \textbf{54.5} & \textbf{58.0} & 56.5 \\
            \bottomrule
        \end{tabular}}        
\end{table}


To further analyze the effect of initialization, we will now examine the impact of transfer learning by pre-training with an alternative X-ray dataset, which may be considered another technique to address the data scarcity issue. The results of this experiment are shown in Table~\ref{tab:result:ini:2}. For this, based on the weights obtained with ImageNet, a pre-training is performed with a dataset from a similar domain (``Pre-trained'' column), for which a larger amount of labeled data is available (in this case, considering 1700 training instances). Then, a fine-tuning process is carried out to the source dataset (``From'' column) and evaluated for the target set (``To'' column). As before, four imbalance levels are assessed, from \textit{H}igh to \textit{N}one. Similarly to the previous table, bold values refer to the best performance, but in this case, they are compared to the best initialization method reported in Table~\ref{tab:result:ini:1}. For example, the value 48.0 in the first row and column \textit{H}igh of Table~\ref{tab:result:ini:2} appears in bold because the best initialization value for this same case in Table~\ref{tab:result:ini:1} is lower ($42.2$). However, the first value in the second column, $47.9$, is not marked because the corresponding one in Table~\ref{tab:result:ini:1} obtains a better result ($61.7$) for weight initialization training.

Knowing this, we can see that transfer learning improves, in general terms, the previous results of the pre-trained models. The parameters of a network trained with data of a similar typology help to find features more suitable to the task at hand. If we pay attention to the \textit{M}edium column, most values are not better than the previous ones. This may happen because the network has to re-learn features from the training set (``From'' dataset), but it has very little positive data, i.e., the minority class is hardening the task of differentiating the classes. In the \textit{H}igh imbalance, however, there is only one sample of positive data that will not affect much in the re-training process. Even though these results are somewhat better, similar to before, it also seems to have convergence issues for the \textit{H}igh imbalance case due to having only one minority sample. 

On the other hand, the figures reported for the \textit{L}ow and \textit{N}one imbalance levels almost outperform every result in the previous experiment, especially in the intra-domain scenarios. Clearly, in the case of a balanced or almost balanced set of data, pre-training with data from a similar typology improves the results as it initializes the network with better parameters that will lead to a better classification. Interestingly, even in inter-domain scenarios, the results, while slightly subdued, remain promising. This suggests that even with domain shifts, transfer learning can provide foundational knowledge that outpaces starting afresh or leveraging broader, less task-specific initializations like ImageNet.

\begin{table}[ht]
    \centering
    \caption{$\mbox{F}_{1}$ results obtained through the transfer learning technique. The initial column specifies the dataset used for model pre-training, the ``From'' column signifies the dataset used for fine-tuning, and the ``To'' column represents the dataset considered for evaluation. The intra-domain cases are underlined in each scenario. Each case is analyzed for four levels of imbalance: \textit{H}igh, \textit{M}edium, \textit{L}ow, and \textit{N}one.}
    \label{tab:result:ini:2}
    \setlength{\tabcolsep}{6pt}
    \resizebox{\textwidth}{!}{\begin{tabular}{lllcccc} 
        \toprule
        \textbf{Pre-trained} & \textbf{From} & \textbf{To} & \textit{H} & \textit{M} & \textit{L} & \textit{N} \\ 
        \midrule
        Pad-BIM 
            & \multirow{3}{*}{Chest-Git} 
                & \uline{Chest-Git}     & \textbf{48.0} & 47.9          & \textbf{78.0} & \textbf{85.9} \\
        BIMCV-COVID 
            &  & Pad-BIM                & \textbf{48.3} & 56.0          & 56.2          & \textbf{67.9} \\
        Pad-BIM 
            &  & BIMCV-COVID            & \textbf{47.9} & 42.2          & \textbf{53.7} & \textbf{53.3} \\ 
        
        \midrule
        
        BIMCV-COVID 
            & \multirow{3}{*}{Pad-BIM} 
                & Chest-Git             & \textbf{47.4} & \textbf{55.7} & \textbf{54.6} & \textbf{56.4} \\
        Chest-Git 
            &  & \uline{Pad-BIM}        & \textbf{58.4} & 61.9          & \textbf{83.3} & \textbf{87.3} \\
        Chest-Git 
            &  & BIMCV-COVID            & \textbf{49.5} & \textbf{52.4} & \textbf{60.9} & \textbf{53.4} \\ 
        
        \midrule
        
        Pad-BIM 
            & \multirow{3}{*}{BIMCV-COVID} 
                & Chest-Git             & 43.7          & 48.0          & 49.6          & 51.5 \\
        Chest-Git 
            &  & Pad-BIM                & 41.9          & \textbf{56.1} & 58.2          & \textbf{58.4} \\
        Chest-Git 
            &  & \uline{BIMCV-COVID}    & 41.9          & 46.1          & \textbf{53.5} & \textbf{57.4} \\ 
        
        \midrule
        
        \multicolumn{3}{r}{\textbf{Inter-domain Avg.}}  & \textbf{46.4} & 51.7 & \textbf{55.5} & \textbf{56.8} \\
        \multicolumn{3}{r}{\textbf{Intra-domain Avg.}}  & \textbf{49.4} & 52.0 & \textbf{71.6} & \textbf{76.9} \\
        \multicolumn{3}{r}{\textbf{Global Avg.}}        & \textbf{47.4} & 51.8 & \textbf{60.9} & \textbf{63.5} \\
        \bottomrule
    \end{tabular}}
\end{table}

\subsection{Data augmentation}

Another approach to address the scarcity of labeled data is to apply transformations to generate synthetic images from the available samples. In this section, the results of this process are analyzed by applying the transformations described in Section~\ref{sec:dataaug}, which include horizontal and vertical shifts, scaling, and rotations. For each of these transformations, the result obtained by increasing the $\alpha$ factor with which they are applied is analyzed. Specifically, the following set of values is considered: $\alpha \in \{0, 1, 5, 10, 15\}$. 

\begin{figure}[ht]
  \centering
  \includegraphics[width=1.0\textwidth]{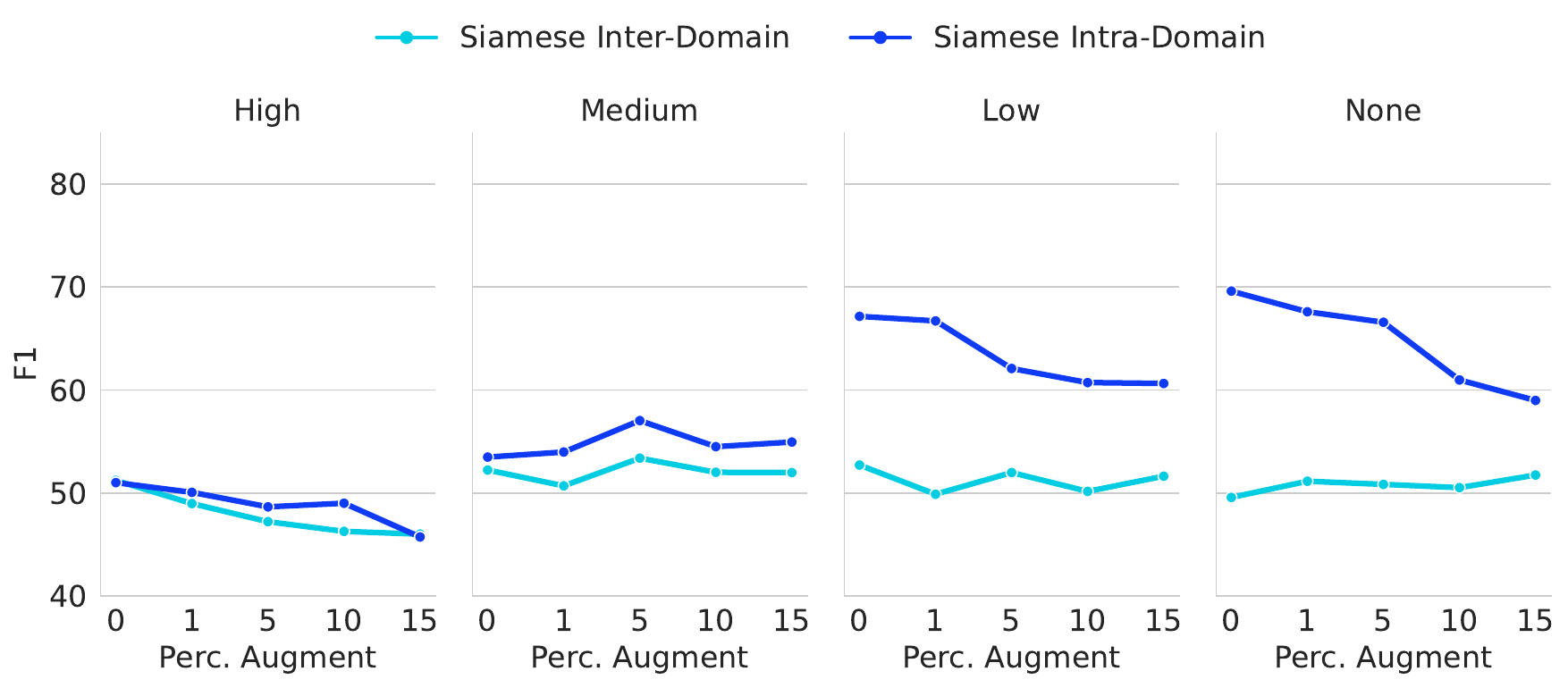}
  \caption{Graph of data augmentation. Five levels of augmented percentage are shown, from $0\%$ to $15\%$, for the four different levels of data imbalance, \textit{H}igh to \textit{N}one.}
  \label{fig:DataAugmentation}
\end{figure}

The graphs depicted in Figure~\ref{fig:DataAugmentation} show the results of these experiments for the four different imbalanced data distributions. In this case, we can see that the trend is of not improving the classification when data augmentation is applied. In some cases, mainly in intra-domain and for high imbalance, data augmentation degrades the performance. This might be caused by the distinctiveness of medical X-ray images. Applying data augmentation includes non-realistic characteristics in the model, hardening the classification process.

\subsection{Dealing with Imbalanced Data}

This set of experiments addresses the data imbalance problem and analyzes the results obtained by applying the techniques proposed in Section~\ref{sec:method:imbalanced}. Table~\ref{tab:results:balance} shows these results, which are similarly arranged as the experiments before, with the training set in ``From'' and the evaluation set in ``To'' columns, respectively. In the table, three cases are evaluated: the weighted loss function (that gives more importance to the minority class, i.e., COVID-19+ cases), the balanced sampling technique by oversampling the minority class to have an equal number of data in the Siamese pairing during the training process, and the combination of both (columns ``Bal. + W.Loss'').

The data in bold refer to the best performance per row and imbalance level. Focusing on the average values at the bottom of the table, we can see that the oversampling technique achieves the best results in \textit{H}igh imbalanced cases. This makes sense as it compensates for the high imbalance by feeding the network with more minority samples. Nevertheless, for the rest of the cases, the average results of combining oversampling with the proposed weighted loss function provide the best classification.  

\begin{table}[ht]
    \centering
    \caption{Comparison of the $\mbox{F}_{1}$ results obtained through the balancing techniques: weighted loss function, oversampling minority data, and the combination of weighted loss and oversampling. Results for the four data distributions considered, from \textit{H}igh to \textit{N}one. The best results per line are marked in bold.}
    \label{tab:results:balance}
    \setlength{\tabcolsep}{2pt}
    \footnotesize
        \resizebox{\textwidth}{!}{\begin{tabular}{lllcccclcccclcccc} 
            \toprule
             &  &  & \multicolumn{4}{c}{Weighted loss}
                &  & \multicolumn{4}{c}{Balanced sampling}
                &  & \multicolumn{4}{c}{Bal. + W.Loss} \\ 
            \cmidrule{4-7}\cmidrule{9-12}\cmidrule{14-17}
            \textbf{From} & \textbf{To} 
                &  & \textit{H} & \textit{M} & \textit{L} & \textit{N} 
                &  & \textit{H} & \textit{M} & \textit{L} & \textit{N} 
                &  & \textit{H} & \textit{M} & \textit{L} & \textit{N} \\ 
            \cmidrule{1-2}\cmidrule{4-7}\cmidrule{9-12}\cmidrule{14-17}

                \multirow{3}{*}{Chest-Git}
                        & \uline{Chest-Git} 
                            &  & 46.5           & 50.4          & \textbf{75.8} & \textbf{76.2} 
                            &  & 49.8           & 51.1          & 71.6          & 74.7 
                            &  & \textbf{50.7}  & \textbf{59.4} & 68.9          & 75.2 \\
                        & Pad-BIM 
                            &  & 51.5           & 54.8          & 56.6          & \textbf{61.7} 
                            &  & \textbf{55.9}  & \textbf{61.5} & 57.0          & 59.9 
                            &  & 45.1           & 56.6          & \textbf{58.8} & 57.5 \\
                        & BIMCV-COVID 
                            &  & 43.9           & 45.5          & \textbf{47.0} & 44.5 
                            &  & \textbf{46.4}  & \textbf{46.9} & 46.9          & 41.0 
                            &  & 44.5           & 46.8          & 43.7          & \textbf{49.2} \\                             
            \cmidrule{1-2}\cmidrule{4-7}\cmidrule{9-12}\cmidrule{14-17}
            
                \multirow{3}{*}{Pad-BIM}  
                    & Chest-Git 
                        &  & 50.3           & 50.0          & 49.0          & 40.1 
                        &  & \textbf{54.4}  & 46.9          & \textbf{49.6} & 38.7 
                        &  & 44.3           & \textbf{51.9} & 42.6          & \textbf{45.9} \\
                    & \uline{Pad-BIM} 
                        &  & 47.9           & 64.3          & \textbf{80.8} & \textbf{82.4} 
                        &  & \textbf{55.0}  & 62.2          & 77.6          & 80.4 
                        &  & 48.9           & \textbf{65.5} & 80.2          & 80.9 \\                    
                    & BIMCV-COVID 
                        &  & 46.4           & 47.7          & 52.8          & \textbf{52.1} 
                        &  & 47.2           & 48.1          & 51.7          & 48.5 
                        &  & \textbf{48.1}  & \textbf{48.7} & \textbf{53.8} & 51.5 \\
                        
            \cmidrule{1-2}\cmidrule{4-7}\cmidrule{9-12}\cmidrule{14-17}

                \multirow{3}{*}{BIMCV-COVID}
                    & Chest-Git 
                        &  & 42.8           & \textbf{55.6} & 49.3          & 50.4 
                        &  & \textbf{51.4}  & 52.8          & 53.0          & 51.7 
                        &  & 48.4           & 53.1          & \textbf{58.0} & \textbf{56.7} \\
                    & Pad-BIM 
                        &  & 48.5           & 55.5          & 51.6          & 55.9 
                        &  & \textbf{52.0}  & \textbf{57.2} & 58.1          & \textbf{57.5} 
                        &  & 45.2           & 52.0          & \textbf{60.3} & 56.5 \\                    
                    & \uline{BIMCV-COVID} 
                        &  & 46.4           & 47.0          & 51.0          & 51.2 
                        &  & \textbf{48.2}  & \textbf{47.2} & \textbf{52.3} & \textbf{53.7} 
                        &  & 45.3           & \textbf{47.2} & 51.9          & 51.9 \\ 
                        
            \cmidrule{1-2}\cmidrule{4-7}\cmidrule{9-12}\cmidrule{14-17}
            \multicolumn{2}{r}{\textbf{Inter-Domain Avg.}} 
                &  & 47.2           & 51.5          & 51.1          & 50.8 
                &  & \textbf{51.2}  & \textbf{52.2} & 52.7          & 49.6 
                &  & 45.9           & 51.5          & \textbf{52.9} & \textbf{52.9} \\
            \multicolumn{2}{r}{\textbf{Intra-Domain Avg.}} 
                &  & 46.9           & 53.9          & \textbf{69.2} & \textbf{69.9} 
                &  & \textbf{51.0}  & 53.5          & 67.1          & 69.6 
                &  & 48.3           & \textbf{57.4} & 67.0          & 69.3 \\
            \multicolumn{2}{r}{\textbf{Global Avg.}} 
                &  & 47.1           & 52.3          & 57.1          & 57.2 
                &  & \textbf{51.1}  & 52.6          & 57.5          & 56.2 
                &  & 46.7           & \textbf{53.5} & \textbf{57.6} & \textbf{58.4} \\
            \bottomrule
        \end{tabular}}
\end{table}

\begin{figure}[ht]
  \centering
  \includegraphics[width=1.0\textwidth]{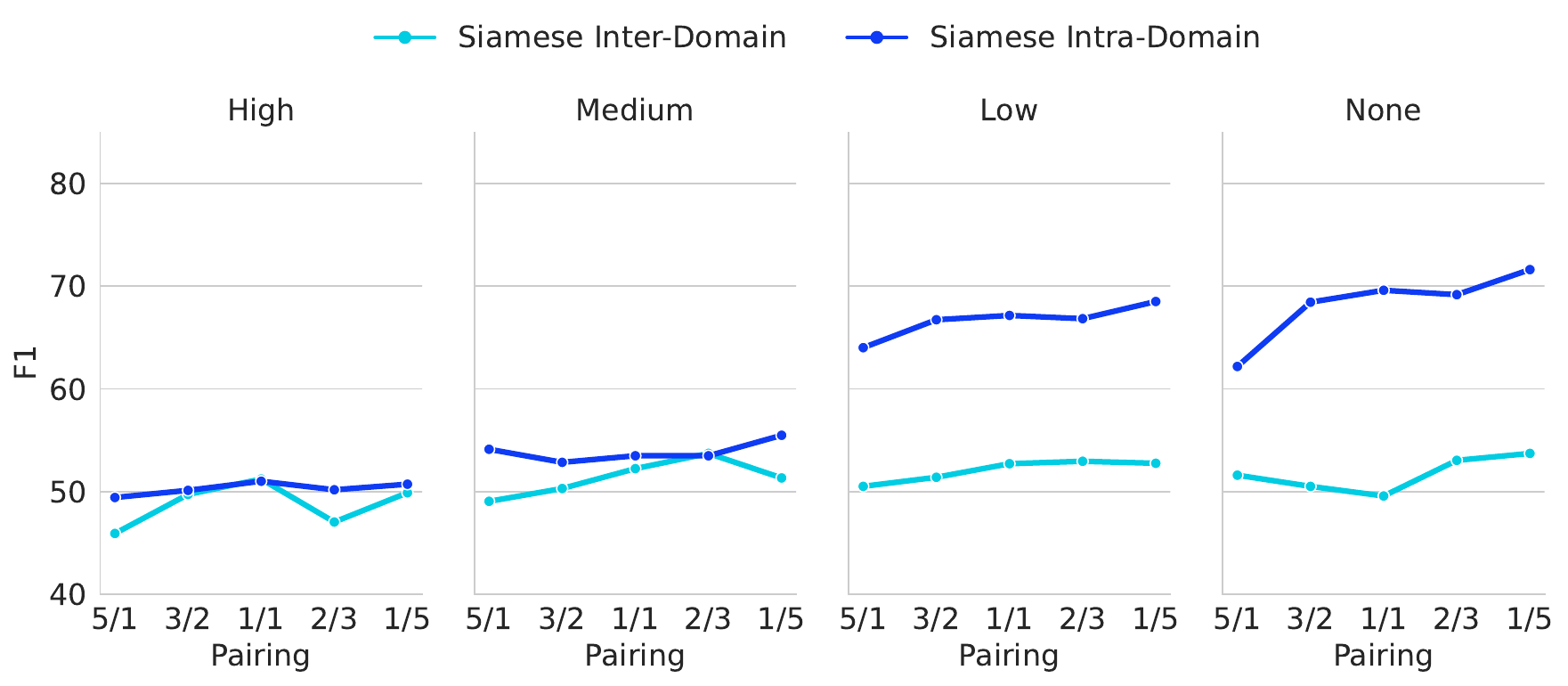}
  \caption{Graph of pairing experimentation. Five different ratios of positive/negative pairs, and \textit{H}igh to \textit{N}one data distribution cases.}
  \label{fig:pairing}
\end{figure}

Next, the fourth proposal to deal with imbalanced data is evaluated: the level of positive/negative data pairing (referring to pairs of equal or different images) during the training process of the Siamese network. Figure~\ref{fig:pairing} presents the $\mbox{F}_{1}$ results for five pairing ratios and for the four data distributions, from \textit{H}igh to \textit{N}one. Particularly, the Siamese network is trained with pairing proportions from five positives for every negative ($5/1$), then three positives for every two negatives ($3/2$), up until one positive for every five negatives ($1/5$). Note that in the case of \textit{H}igh imbalance, where there is only one positive sample (COVID-19 infected patients) along with 100 negative (healthy) data, the $5/1$ pairing will only have this same image for the negative pairs. Consequently, this image will be presented to the network in every batch, leading to overfitting. Therefore, when evaluated with varied positive data (other COVID cases), the classification accuracy will drop. This phenomenon is further analyzed in the following paragraph.

From Figure~\ref{fig:pairing}, we can observe that in the high imbalance scenario, the pairing hardly affects the performance, obtaining results around 50 of $\mbox{F}_{1}$ in all the pairing levels studied, which denotes the previously mentioned problem: the sparse positive data (COVID cases) in the dataset leads the network to overfit and underperform on the test set. However, in \textit{L}ow and \textit{N}one imbalance, the intra-domain $\mbox{F}_{1}$ is notably higher and improves as more negative pairs are presented to the network. The reason behind this is that when more negative pairs (i.e., different) are fed to the Siamese network, it learns better features to distinguish the classes and, hence, classifies better.  

As a summary of this approach to handling imbalanced data, we can conclude that adjusting the pairing level has no effect in situations with high imbalance. In the cases of similar distributions, using pairing levels with a greater number of negative pairs seems beneficial for the Siamese training process. 

\subsection{Inference classifier}

This section focuses on analyzing the improvement provided by the final classifier used in the proposed pipeline. We have previously reported the results using the histogram method (see Section~\ref{sec:method:inference})---simply choosing the class that minimizes the distance---as it represents the commonly used approach. These results are now compared with those obtained using three alternative classifiers trained from the embeddings generated by the Siamese network, namely $k$NN, Rf, and SVM, and using the same distance metric as before, that is, the Euclidean distance. For each of these methods, their hyperparameters were initially studied, eventually selecting the best configurations, which include $k$ values within the range $k \in [1, 15]$, a number of tree estimators $t \in [10, 500]$ for Rf, and Linear, Polynomial, and Radial Basis functions for the kernel of SVM with a learning cost $c \in [1, 9]$. 

Table~\ref{tab:results:inferenceClassifier} shows the outcomes of these experiments, comparing the performance of the four classifiers across inter- and intra-domain levels and for the four imbalanced distributions considered. From a general analysis of these results, it is observed that the SVM classifier reports an improvement in all scenarios except for high levels of imbalance, for which the use of the histogram-based or kNN-based approaches seems to be more advisable. If we analyze the results at the inter- and intra-domain levels, it is observed that SVM generates a model that generalizes better to other domains, while the solutions based on histogram and $k$NN are more effective within the same domain.

\begin{table}[ht]
    \centering
    \caption{Comparison of the $\mbox{F}_{1}$ results obtained by the different inference classifiers considered: Histogram, $k$NN, Rf, and SVM. The best result for each imbalanced scenario is marked in bold.}
    \label{tab:results:inferenceClassifier}
    \setlength{\tabcolsep}{6pt} 
    \footnotesize
        \begin{tabular}{lclcccc} 
            \toprule
             & \textbf{Imbalance level} 
                &   & \textit{Hist} 
                    & \textit{kNN} 
                    & \textit{Rf} 
                    & \textit{SVM}  \\ 
            \cmidrule{1-2}\cmidrule{4-7}
            \multirow{4}{*}{\textbf{Inter-domain}} 
                & \textit{High}     &  & 45.4           & \textbf{49.1} & 35.1 & 45.6  \\
                & \textit{Medium}   &  & 47.7           & 50.6          & 40.6 & \textbf{51.8}  \\
                & \textit{Low}      &  & 49.3           & 51.5          & 53.6 & \textbf{54.3}  \\
                & \textit{None}     &  & 54.4           & 54.4          & 58.4 & \textbf{56.9}  \\ 
            \cmidrule{1-2}\cmidrule{4-7}
            \multirow{4}{*}{\textbf{Intra-domain}} 
                & \textit{High}     &  & \textbf{51.0}  & 47.4          & 34.7 & 42.1  \\
                & \textit{Medium}   &  & 53.1           & 52.7          & 42.1 & \textbf{54.1}  \\
                & \textit{Low}      &  & 64.7           & 65.9          & 65.7 & \textbf{67.1}  \\
                & \textit{None}     &  & 71.8           & 71.7          & 71.7 & \textbf{72.2}  \\ 
            \cmidrule{1-2}\cmidrule{4-7}
            \multicolumn{2}{r}{\textbf{Inter-domain avg.}} 
                                    &  & 49.2           & 51.4          & 46.9 & \textbf{52.1} \\ 
            \cmidrule{1-2}\cmidrule{4-7}
            \multicolumn{2}{r}{\textbf{Intra-domain avg.}} 
                                    &  & \textbf{60.2} & 59.5           & 53.5 & 58.9  \\ 
            \cmidrule{1-2}\cmidrule{4-7}
            \multicolumn{2}{r}{\textbf{Global Avg.}} 
                                    &  & 52.8           & 54.1          & 49.1 & \textbf{54.4} \\
            \bottomrule
        \end{tabular}
\end{table}

\subsection{Analysis of the training set size}

In this section, the performance of the proposal is evaluated as the training set size increases. These results are also compared with those obtained by training a single backbone (that is, the CNN ResNet-50 v2 architecture, which is also the one analyzed in the previous work~\cite{ibpria}). Regarding the size of the training set, in addition to the data distributions with 100 samples for the majority class, which has been used in the previous experiments, the amount of data is increased to 200 and 300 samples following the same imbalanced distributions: \textit{H}igh $\rightarrow \{100/1, ~200/2, ~300/3\}$, \textit{M}edium $\rightarrow \{100/10, ~200/20, ~300/30\}$, \textit{L}ow $\rightarrow \{100/50, ~200/100, ~300/150\}$, and \textit{N}one $\rightarrow \{100/100, ~200/200, ~300/300\}$. 

\begin{figure}[ht]
  \centering
  \includegraphics[width=1.0\textwidth]{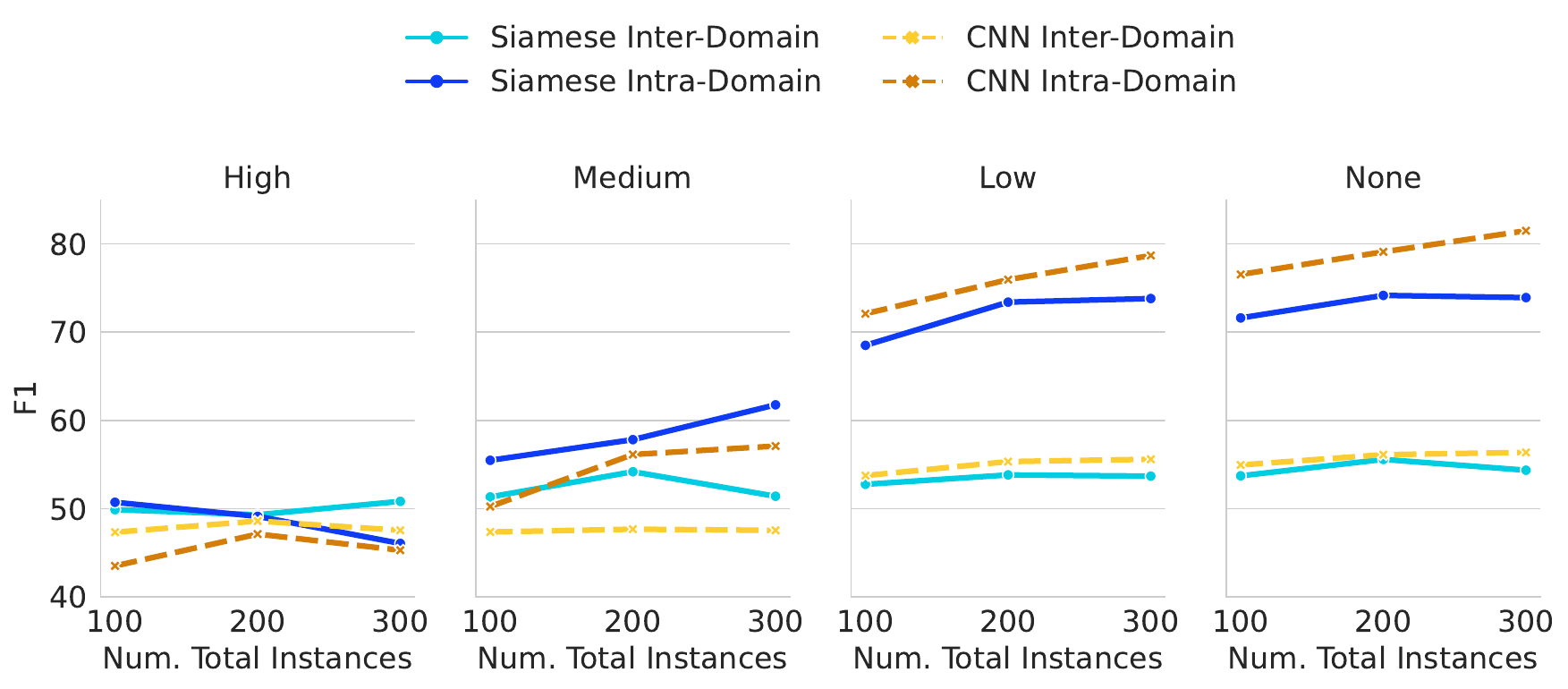}
  \caption{Graph comparison of Siamese and CNN network architectures. The evaluation is carried out for three sizes of training sets, 100, 200, and 300 samples for the majority class, and \textit{H}igh to \textit{N}one imbalance data distributions.}
  \label{fig:trainSizeComp}
\end{figure}

The results of these experiments are depicted in Figure~\ref{fig:trainSizeComp} for both the Siamese network and the CNN at the inter- and intra-domain levels. The first aspect to highlight is that in the case of \textit{H}igh imbalance, the error is quite similar for both models, achieving a low $\mbox{F}_{1}$ performance and being the CNN the lowest in most cases. This shows that the two architectures have problems learning this highly imbalanced distribution.

Generally, the lower the imbalance, the better the results for the intra-domain scenarios. When studying the \textit{L}ow and \textit{N}one cases, we can see that intra-domain models are remarkably better, being the CNN slightly better in both cases. An additional observation from the graphs is that the Siamese network stabilizes earlier than the CNN. 

From the information in the charts of Figure \ref{fig:trainSizeComp}, we can conclude that the Siamese network works better for \textit{H}igh and \textit{M}edium-imbalanced datasets. In contrast, using this network is not necessary in cases of balanced data. On the other hand, the fact that the inter-domain training processes maintain a low $\mbox{F}_{1}$ score regardless of the imbalance level demonstrates that the networks do not generalize properly.

\subsection{Discussion}

This last section summarizes the improvements provided by each technique studied for a few-shot learning scenario with imbalanced datasets. These results are compared with those obtained in the previous study~\cite{ibpria} using equivalent techniques for imbalanced datasets but applied to a CNN when there is no limitation of labeled data. This comparison aims to shed light on whether these techniques are consistent in their results or, on the contrary, performance depends on the amount of information available or the network architecture.

Table~\ref{tab:results-summary} shows a summary of results for all the previous approaches and both inter- and intra-domain cases, indicating the percentage of improvement relative to the base case, which is the model trained from scratch as described in Section~\ref{sec:initialization}. For the sake of fair comparison, the percentages of improvement shown of the Siamese network are with \textit{M}edium imbalance (100/10) since it represents the data distribution most similar to the original one studied in the previous work using a CNN. In the table, the CNN cases without results are marked with ``--'', either because they were not considered in the previous study or because they do not apply to a CNN, such as the level of pairing.

From this general analysis, it can be observed that the various techniques studied offer promising improvements over the baseline in almost all cases. However, it appears that using one technique over another may be more advisable depending on whether the learning problem involves limited data or if there are no restrictions on labeled data. In the case of Few-shot learning, it seems more advisable to have a better initialization and use a classifier learned from the embeddings of the Siamese network during inference. However, if there are no data restrictions, using oversampling and weight loss proves to be much more beneficial. This may be because, in a Few-shot scenario with a single sample, if repeated many times or given a high weight, it generates overfitting towards the minority class, limiting its generalization capabilities.

The best technique to select also depends on the application domain. For instance, in few-shot scenarios, if the goal is to have better inter-domain generalization, the use of transfer learning and SVM is recommended. On the other hand, if the aim is to be more effective within the source domain, a general initialization---with ImageNet, which does not create a bias towards other distributions---is more appropriate, employing the proposal for oversampling combined with a weighted loss function. When there are no data restrictions, these conclusions change slightly. For example, in addition to weight loss and oversampling---which in this case are recommended to be used separately since they provide a more notable improvement---it is always advisable to initialize using transfer learning. This may be because having more data available for fine-tuning eliminates the risk of creating bias.

\begin{table}[ht]
    \caption{Summary of the improvements obtained by each of the techniques proposed for the Siamese architecture (in the case of the inference classifier, only the two best are included). These results are compared to those obtained using equivalent techniques on a CNN in our previous work~\cite{ibpria}.}
    \label{tab:results-summary}
    \setlength{\tabcolsep}{2pt}
    \footnotesize
    \resizebox{\textwidth}{!}{\begin{tabular}{llcclcccccccccc} 
        \toprule
            & & \multicolumn{2}{c}{\textbf{Initialization}}                   &  & \multicolumn{2}{c}{\textbf{Inf. Classifiers}} &  & \textbf{Data Aug.} &  & \multicolumn{5}{c}{\textbf{Imbalance solutions}} \\ 
        \cmidrule{3-4}\cmidrule{6-7}\cmidrule{9-9}\cmidrule{11-15}
        
        & & \rotatebox{90}{\textbf{ImageNet}} & \rotatebox{90}{\textbf{Tr. learning}} & & \rotatebox{90}{\textbf{$k$NN}} & \rotatebox{90}{\textbf{SVM}} &  & \rotatebox{90}{\begin{tabular}[l]{l}\textbf{ImageNet} \\ \textbf{+ Data Aug.}\end{tabular}} &  & \rotatebox{90}{\textbf{Oversampling}} & \rotatebox{90}{\textbf{W. Loss}} & \rotatebox{90}{\begin{tabular}[l]{l}\textbf{Oversampling} \\ \textbf{+ W. Loss}\end{tabular}} &  & \rotatebox{90}{\begin{tabular}[l]{l}\textbf{Oversampling} \\ \textbf{+ Pairing}\end{tabular}}  \\ 

        \midrule
        \multirow{3}{*}{\textbf{CNN}} & \textbf{Inter-Domain} 
            & 2.1\% & 2.7\% & & - & - &  & 2.8\%  &  & 8.1\% & 10.2\%  & - &  & - \\
        & \textbf{Intra-Domain} 
            & 2.0\% & 3.8\% & & - & - &  & 1.9\% &  & 5.6\% & 8.9\% & - & & -\\ 
        \cmidrule{2-15}
        & \textbf{Average} 
            & 2.1\% & 3.1\% & & - & - &  & 2.5\% &  & 7.3\% & 9.8\% & - &  & - \\ 
        \midrule
        \multirow{3}{*}{\textbf{Siamese}} & \textbf{Inter-Domain} 
            & 2.7\% & 4.0\% & & 2.9\% & 4.1\% &  & -0.2\% &  & 1.7\% & 1.0\% & 1.0\% &  & 0.8\% \\
        & \textbf{Intra-Domain} 
            & 5.4\% & 0.6\% & & -0.4\% & 1.0\% &  & 3.9\% &  & 1.8\% & 2.2\% & 5.6\% &  & 3.7\% \\ 
        \cmidrule{2-15}
        & \textbf{Average}      
            & 3.6\% & 1.9\% & & 1.8\% & 3.1\% &  & 1.1\% &  & 1.7\% & 1.4\% & 2.5\% &  & 1.8\% \\
        \bottomrule
    \end{tabular}}
\end{table}

\section{Conclusions}
\label{sec:conclusions}

This study delves into the performance of various techniques in the challenging context of few-shot learning with imbalanced medical datasets. The results shed light on the intricate dynamics between the amount of data, distribution imbalance, and model architecture. While some of the studied techniques are well-established in the literature, others are not, such as the adaptation proposals to deal with imbalanced data. Besides, this work focuses on evaluating their effectiveness in the context of medical imaging and examining their performance when used in combination with Siamese architectures. 

First, we focus on the initialization of network parameters for few-shot scenarios. The main conclusion is that pre-training the model using transfer learning, either with general data or with data from a similar domain, helps improve the generalization capabilities of the model in this challenging data-sparse scenario. Several data augmentation techniques have also been studied, concluding that applying standard transformations with medical imaging for few-shot scenarios is not a good practice due to the peculiarities of these data. 

Furthermore, four approaches have been proposed to address data imbalance, including a weighted loss biased to the minority class, balancing the samples, and modifying the pairing ratio of positive and negative samples. The conclusions are that, in cases of high imbalance, balancing the samples by repeating the minority data helps improve the results. However, when the dataset is not highly imbalanced, combining a weighted loss with balanced data allows the network to learn better features. Different pairing ratios between the same and other classes in the Siamese training were also studied. In this case, when the datasets are balanced and the pairing ratio shifts towards more negative (different) pairs than positive, intra-domain results improve since this helps to learn features that distinguish between classes.

Regarding classification, the evaluation included four approaches: Histogram, $k$NN, Rf, and SVM. The SVM classifier is more accurate in all inter- and intra-domain scenarios except for high levels of imbalance. In high imbalance, using the histogram-based (for intra-domain) or kNN-based (for inter-domain) approaches reports better results.

Finally, we compared the Siamese network (with the different techniques introduced for dealing with few-show and imbalanced datasets) against a standard CNN network from previous works. We first studied the impact of the training set with other data distributions. The main conclusion of this experiment is that, in highly imbalanced situations, the performance of both Siamese and standard CNN is low, with the first slightly better. However, in balanced cases, the inter-domain training improves with the dataset size, whereas the inter-domain does not, showing limited generalization capabilities. Afterward, we compared the different initializations, data augmentation, and imbalance solutions for the CNN and the Siamese network. As expected, the general observation from this study validates the general intuition that the specific technique to be applied depends on the level of data available and the application domain.

For future work, these techniques could also be adapted and studied for matching, prototypical, and relation networks to compare them to the Siamese approach. In addition, alternative network architectures other than ResNet-50 could be evaluated. Data augmentation guided by experts for the medical domain could also be included, as well as additional datasets. Regarding initialization, alternative techniques, such as Self-Supervised Learning, could be evaluated for scenarios with data scarcity.

\section*{Statements and Declarations}

\noindent
\textbf{Funding} No external funding was received to carry out this research.\\

\noindent
\textbf{Competing interest} The authors have no relevant financial or non-financial interests to disclose.\\

\noindent
\textbf{Ethics approval} For this article, the authors did not undertake work that involved humans or animals.\\

\noindent
\textbf{Data availability} All datasets are publicly accessible: ChestX-ray can be found at \url{https://nihcc.app.box.com/v/ChestXray-NIHCC}, GitHub-COVID at \url{https://github.com/ieee8023/covid-chestxray-dataset}, PadChest is available at \url{https://bimcv.cipf.es/bimcv-projects/padchest}, and BIMCV-COVID repositories can be accessed through \url{https://bimcv.cipf.es/bimcv-projects/bimcv-covid19}. The source code is available upon request to the corresponding author.\\

\noindent
\textbf{Authors’ contributions} All authors contributed to the study's conception and design. A.G.C. performed material and methods preparation and carried out the experimentation. All authors analyzed the results and contributed to the writing of the final manuscript.


\bibliographystyle{elsarticle-num}
\bibliography{Arxiv/refs}

\end{document}